# Myocardial Contractions and the Ventricular Pressure–Volume Relationship

Paola Nardinocchi * Luciano Teresi † Valerio Varano ‡

May 28, 2010

## Abstract

We present a reduced–order heart model with the aim of introducing a novel point of view in the interpretation of the pressure–volume loops. The novelty of the approach is based on the definition of active contraction as opposed to that of active stress. The consequences of the assumption are discussed with reference to a specific pressure-volume loop characteristic of a normal human patient.

## 1 Introduction

The heart is a specialised muscle that contracts regularly and pumps blood to the body and the lungs. The centre of the pumping function are the ventricles. Due to the higher pressures involved, the left ventricle is especially studied. The pumping action is caused by a flow of activation potential through the heart that repeats itself in a cycle. The effectiveness of the pumping action may be evaluated through the analysis of different parameters: the stroke volume, the ejection fraction, the end–systolic pressure–volume relationship, etc.... All of these are well represented in the pressure–volume loop which, in the end, represents an assessment of the overall mechanical activity of the heart; the reader is referred to the excellent review (1), or to (2).

Ventricular chamber pressures and volumes are related through the contractile behaviour of the muscular structure of the cardiac tissue. A qualitative and quantitative bridge between the function of the heart as a whole and the microscopic dynamics characterizing muscle contraction would be desirable but it is still unrealized (3). There exists complex mathematical models relating ventricular chamber pressure and volume to muscle length and force generation which are the basic notions characterizing the muscle function at a macroscopic level. Nevertheless, it is well known that reduced–order heart models would be helpful in a qualitative assessment of the global state of health of the left chamber (4).

Here, we present and discuss a novel point of view in the interpretation of the pressure–volume loops based on a special modeling of muscle tissues and aimed to introduce new insights into the mechanisms determining the performance of the ventricular chamber.

The first issue is the modeling of the contractile behaviour of the muscular structure of the cardiac tissue. From the microscopic point of view, the excitation-contraction coupling (ECC) in cardiac muscular fibres is a complex mechanism involving many variables such as membrane potential, ionic conductance, intracellular calcium concentration, membrane strain and stress, and changes in the rest length of muscle fibres due to the interaction of actin and myosin. The definition of a realistic model that is sufficiently complete to account for the principal mechanism involved in ECC, yet simple enough to be effective, is an arduous task. The issue has been addressed from the microscopic point of view in various papers (5), (6), (7). However, the muscle fibres ability to contract and relax (in response to biochemical signals) involves the function of the heart as a whole. Hence, it is mandatory to have a macroscopic model of the contractile properties of cardiac tissues in order to discuss the pumping function of the heart in terms of contraction as well as pressure and volume.

*Università di Roma "La Sapienza", Roma, Italy, paola.nardinocchi@uniroma1.it

†Università Roma Tre, Rome, Italy, teresi@uniroma3.it

‡Università Roma Tre, Rome, Italy, varano@uniroma3.it

Our modeling of muscle tissues accounts for muscle contraction through the notion of active deformation. We assume that the contraction experienced by a fibre of muscle tissue under stimulus is described at the macroscopic scale by a (stress–free) change, called active deformation, in the length of the fibre. We say that a muscle fibre stays in its *active* state when it is contracted; otherwise, it stays *slack*. It is worth noting that, from a mechanical point of view, the active state of a muscle fibre is a ground state as determined by a change of length due to electrochemical stimuli. The *visible* state of the muscle fibre is determined from the active state by the amount of stress the fibre sustains. Roughly speaking, the active deformation describes how muscle tissues shorten once activated and are left free to contract, while the visible deformation describes the state that a muscle tissue attains once contracted, loaded and/or kinematically constrained (as in isometric activation). The material model based on our modeling of muscle tissue is called the *two-layer* model (8), (9), (10), (11). The visible layer, which is observable and measurable also *in vivo*, shows muscle fibres while working; the hidden layer shows muscle fibres in their excited state before working. We used this material model to study the mechanics of the left ventricle following a top–down modeling approach as in (12).

Within the context of our special modeling, PV loops may be looked at in a novel way which may introduce new insights into the mechanism determining the performance of the ventricular chamber. Indeed, PV loops are derived from pressure and volume measures performed during the cardiac cycle; we could say that they are pictures of the visible state of the heart. Our modeling, with its special notion of contraction, defines a correspondence between each point of a PV loop and the tissue contraction which determined it. In the end, a PV loop may be viewed as determined by three main state variables, pressure, volume, and contraction; changes in ventricular function visualized through PV loops may be easily interpreted in terms of changes in these variables.

Severe geometrical models of left ventricle are used to estimate the overall mechanical activity of the heart. We refer to a spherical approximation of the left ventricle: the key mechanical relationships are not markedly altered by the simplified geometry (1) and we can easily present and discuss our different point of view (see (12) for more complex conceptual geometries). Moreover, we do not account for any tissue anisotropy: both the passive and the active response of the tissue are assumed to be completely isotropic. Nevertheless, the large deformations of the ventricle make it mandatory to set the modeling within the context of non–linear elasticity.

To better illustrate and discuss our point of view, we use a specific sampling of pressure–volume pairs measured by (15) with reference to a normal human patient. We use the corresponding PV loop, shown in figure 6, to discuss the main characteristics of our reduced–order heart model.

Let us note that here we are not interested in relating heart electrophysiology to myocardial contraction; a proposal about this issue, based on the notion of active contraction, as been presented in (13), (14).

# 2 The pressure–volume loops

Of the four chambers that comprise the whole heart, the left ventricle (LV) accomplishes the major mechanical work, while undergoing large deformations and intense stress states. On a simplistic level, the ventricle is an ellipsoidal chamber, whose walls are composed of muscle fibres. It is the *contraction* originated in the muscles that translates into pressure and/or volume changes of the chamber.

The LV cycle may be schematized as the sequence of four steps: filling–the *diastolic* phase; isovolumetric contraction; ejection–the *systolic* phase; isovolu-metric relaxation. During the cycle, both pressure and volume vary in time, and a quite useful determinant of the cardiac performance is the plot representing the pressure-volume relationship in the LV during the entire cycle, the so-called PV loop; some of the many clues contained in the plot (see Fig. 1) are briefly summarized in the following.

Point 1 defines the end of the diastolic phase and is characterized by the end–diastolic volume $v_{ed}$ (EDV) and pressure $p_{ed}$ (EDP). At this point the mitral valve closes and cardiac muscle starts to *contract* in order to increase the blood pressure. At point 2 the systolic phase begins: the aortic valve opens and blood is ejected outside the LV. Muscles continue contracting in order to further the ejection, while volume decreases to a minimum. Point 3 defines the end of the systolic phase, and is characterized by the end–systolic volume $v_{es}$ (ESV) and pressure $p_{es}$ (ESP). Starting from here, LV undergoes an isovolumic *relaxation* until point 4, where mitral valve opens and filling begins. During the filling phase, muscles continue relaxing in order to accomodate a large increase in blood volume, while maintaining the pressure at a quite low level. Filling is completed at point 1. The difference between maximum and minimum volume is called stroke volume: $v_{str} = v_{ed} - v_{es}$.

Two important curves are usually represented in a PV diagram: the end–diastolic pressure–volume relationship (EDPVR) and the end–systolic pressure–volume relationship (ESPVR), see Fig. 2. These curves characterize the *passive* mechanical response of the ventricle in two quite different states: the relaxed state, and the contracted one, respectively.

Let us consider point 1: muscles are in their most relaxed state (the slack state), and any pressure variation will cause a volume change along the EDPVR, *provided the muscles stay inactive*. Thus, the EDPVR provides a lower boundary for the pressure at which the mitral valve closes, and the position of point 1 depends on the end-diastolic filling pressure. Physiologically, the EDPVR changes as the heart grows during childhood; most other changes accompany pathologic situations (hypertrophy, infarct, dilated cardiomyopathy).

Let us now consider point 3: muscles are in a highly activated state, and the LV behaves as a much stiffer chamber; any pressure variation will cause a volume change along the ESPVR, *provided muscles maintain the same level of activation*. Thus, the ESPVR provides an upper boundary for the pressure at which the aortic valve closes, and the position of point 3 depends on the end-systolic ejection pressure or volume.

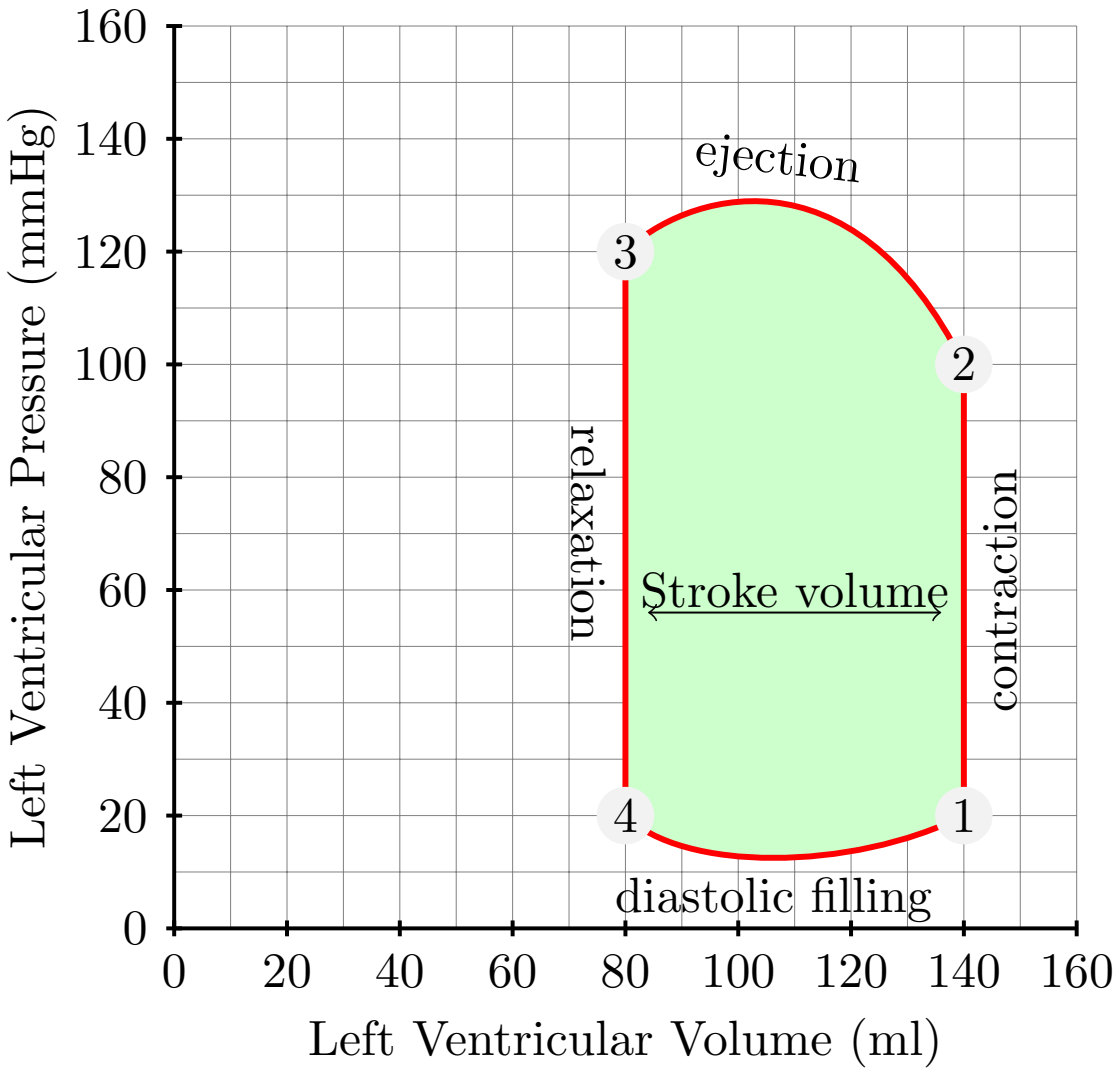

Figure 1: Phases of the cardiac cycle of a normal human patient. 1) Mitral valve closes; isovolumetric contraction. 2) Aortic valve opens; ejection. 3) Aortic valve closes; relaxation. 4) Mitral valve opens, filling. The green area represents the stroke work.

As it is well known, the *preload* denotes the inflow conditions determined by the venous system, and the *afterload*, the outflow conditions determined by the arterial system. The interaction between the LV, the preload, and the afterload strongly influence the arterial blood pressure and the cardiac output, which in turn constitute two key factors for the assessment of the overall cardiovascular performance. Thus, a change in preload or in afterload conditions may markedly alter the PV loop, and results in a shift of point 1 or point 3 along the EDPVR or the ESPVR, respectively, as figure 2 shows.

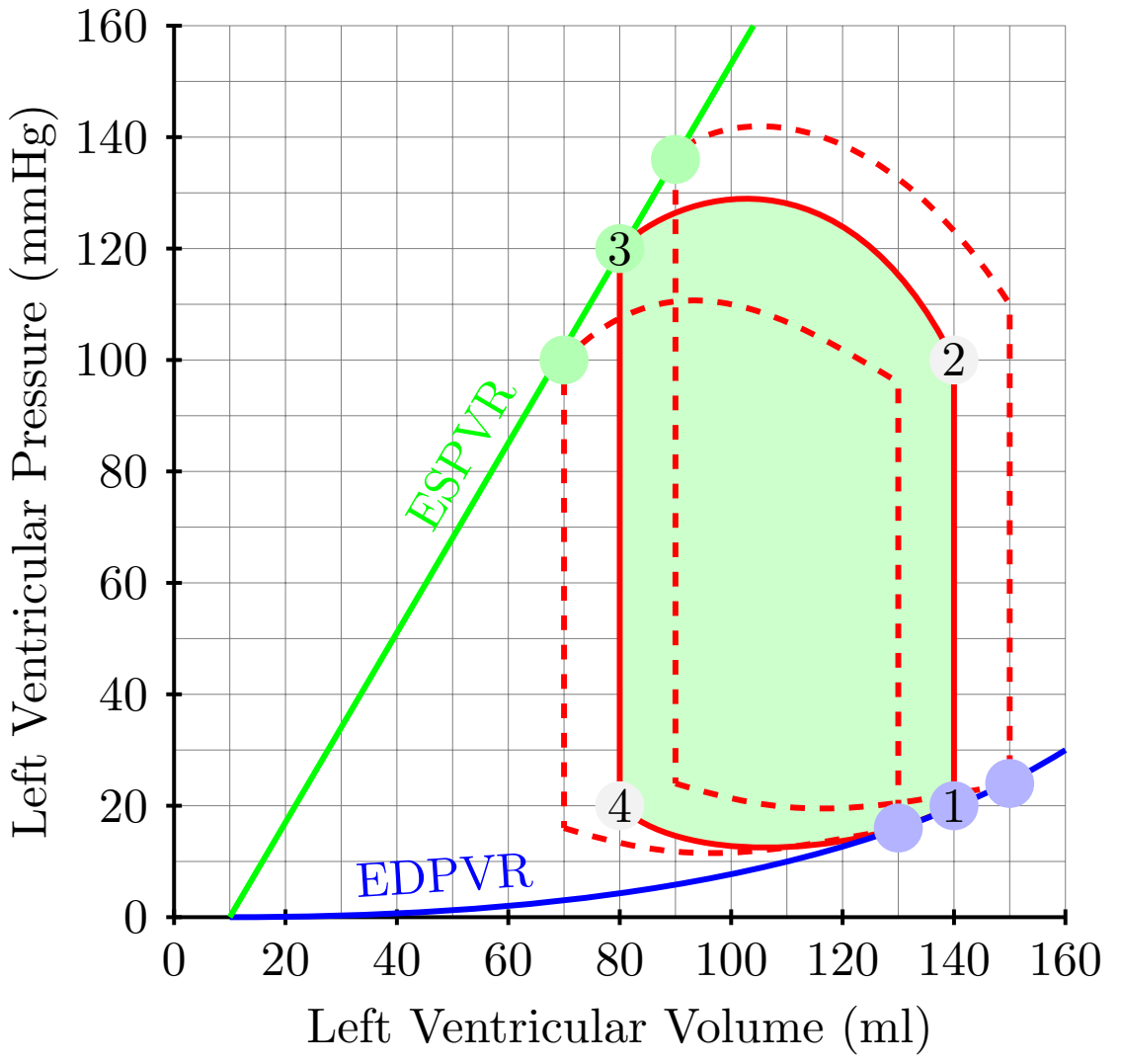

Figure 2: EDPVR and ESPVR represent the pressure-volume relationships for a complete relaxed state, and a highly activated state, respectively. Changes in EDP make point 1 move along the EDPVR; changes in ESP move point 3 along ESPVR.

A meaningful measure of the preload would probably be the sarcomeres length at end diastole. Due to the intrinsic difficulties related to this measurement, EDP is the most common index of preload. Afterload is in general related to the arterial system, but also pathologic conditions, such as a leaky mitral valve or a stenotic aortic valve, could be accounted for.

At any point in the PV loop there corresponds a specific level of muscle activation, and thus, a specific average sarcomere length. A quantitative sampling of sarcomere lengths versus volume is depicted in figure 3. At the preload condition sarcomeres are slightly stretched but still not activated, then passive elasticity predominates. Isovolumic contraction and ejection are the consequences of increasing muscle activation, and thus, an increasing sarcomeres shortening. During isovolumic relaxation and subsequent filling, sarcomeres elongate, and eventually recover their initial length.

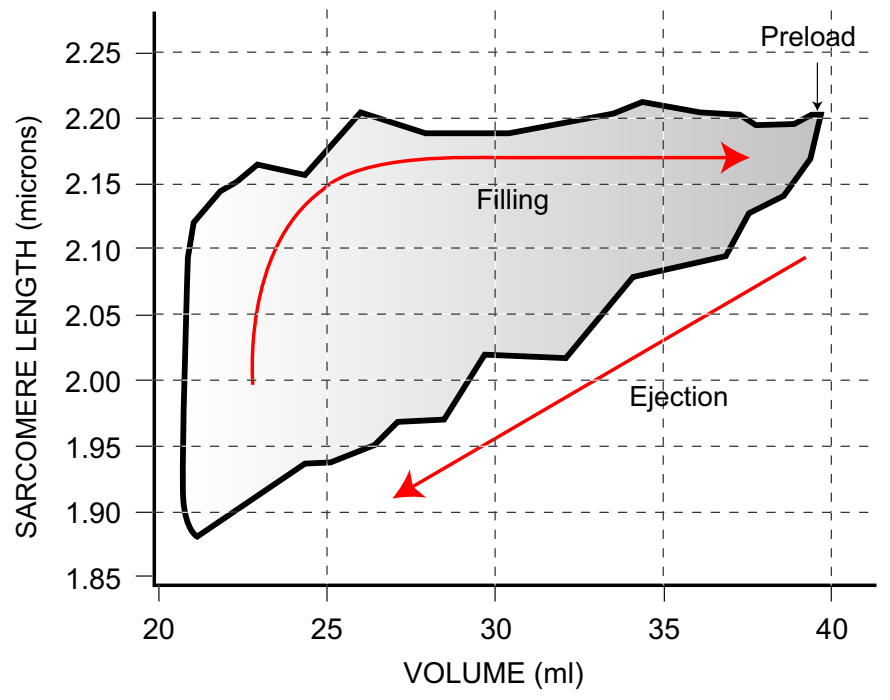

Figure 3: Sarcomere length as function of volume (Elaboration from (16).

## 2.1 Elastance, compliance, and contractility

*Elastance* is defined as the ratio of pressure change to volume change, provided the level of muscles activation remains the same. The notion of elastance is thus introduced to describe the variations of the apparent stiffness of the chamber during the cycle, and it is related to the actual level of activation.

For example, let us consider again the two curves EDPVR and ESPVR. Both curves correspond to a constant level of muscle activation, zero, and very high, respectively. Along the first curve elastance is clearly increasing; along the second one, it is almost constant. *Compliance* is simply the inverse of elastance, that is, the ratio of volume change to pressure change. Thus, we could say that EDPVR is very compliant at low pressure and very stiff at high pressure.

The notion of time–varying elastance $E(t)$ has been introduced to describe the pressure-volume relationship during the entire cycle: denoted with $v_o$ a reference volume corresponding to zero pressure, we have

$$p(t) = E(t)(v(t) - v_o)\,. \qquad (2.1)$$

The idea underlying (2.1) is that of a state function linking three fundamental variables: for example, the knowledge of the time course of elastance $E(t)$ and volume $v(t)$ during the cycle allows for the predic-tion of the time course of pressure. The notion of elastance, introduced by Suga in the late 60s, has been a key concept for cardiac physiologists and cardiologists since then; see (17) for review.

The notion of contractility is strictly related to that of elastance. Indeed, contractility alters the apparent stiffness of muscles. For example, when positive or negative inotropic agents are administered to the heart, the end systolic elastance $E_{es}$ increases or decreases, respectively. For this reason, $E_{es}$ is considered to be an index of contractility. Actually, in the clinical arena, due to the difficulty in measuring the ventricular volume and consequently in assessing $E_{es}$, the ejection fraction (EF), defined as $EF = v_{str}/v_{ed}$, is preferred as an index of contractility.

### 2.2 Pressure–volume loops and diseases

Each pressure–volume loop is determined by the properties of the heart by the characteristics of the circulatory system and by the interplay between the heart and circulation. Pathological circumstances can act on any of these elements and alter the corresponding PV loop with respect to the normal status. In (2), different pictures show the changes in EDPVR, ESPVR, and PV loop under decreased and increased contractility, increased end–diastolic volume, vasoconstriction, fluid retention, and increased lusitropy (that is an index of filling)[1]

Obviously, there is not a one–to–one correspondence between cardiac pathologies and altered PV loops. As it is explained in (2), a shift of the ESPVR downward and to the right may correspond to an infarcted myocardium as well as to a dilated cardiomyopathy. Both the pathologies are characterized by decreased contractility which, ultimately, determines the transformation of the PV loop.

The neurohumoral response, which has the role of performing within our bodies adaptation to changes in the external world, may operate hemodynamical adjustments identifiable by alterations of the PV loop. The right therapies able to reverse these alterations may have positive and negative effects on the pumping function of the heart. Still, it is difficult to make a one–to–one correspondence between a therapy and its effect on the PV cycle of the heart.

In general, heart failure corresponds to a number of abnormalities in most of the structures involved in the excitation–contraction mechanism which is at the basis of the pumping function of the heart. Such abnormalities result from alterations of the shape, the size, and the composition of the heart. As an example, abnormal cardiac performance in patients with chronic heart failure are caused by eccentric and concentric hypertrophy which are revealed by alterations of pressure–volume loops determined by different causes acting simultaneously.

In the end, it seems that the interdependence of the key variables determining pressure–volume loops makes it difficult to establish a direct correspondence between an alteration of the pumping function of the heart and the direct pathology without further evidence of the history of the patient, the electrophysiological state of the tissue and other information which PV loops do not show.

## 3 Modeling of muscle tissues

The motor units of muscles, *sarcomeres*, exist at a microscopic scale. A sarcomere contains many proteinaceous filaments, the actin and the myosin. The interaction of these filaments through cross-bridges is responsible of the contractile properties of the muscle. Upon activation of the muscle, the actin and myosin filaments move with respect to each other causing the sarcomere to shorten. Upon de-activation of the muscle the actin and myosin filaments can recover their original positions, pulled back by elasticity in the surrounding tissue.

Here, we discuss a macroscopic model of muscle which embodies the notion of contraction (10), (12). This model introduces in the realm of biomechanic ideas developed long ago to describe some phenomena related to the change of ground states in elastic bodies such as plasticity or phase transitions (18).

We suppose that muscle fibres are tightly embedded in the tissue and perform a sort of homogenization through the identification of a fibre of muscle tissue with a muscle fibre. The key issue is to distinguish between active and passive deformations of the fibre. The active deformation describes the contraction of an activated and unloaded muscle fibre, that is, the shortening the fibre would have if left free to contract. The passive deformation measures the difference between the unloaded, contracted state, and the visible state which is, in vivo, the sole observable state. It is worth noting that this point of view considers as primary the notion of contraction, instead of that of tension: muscle tension arises whenever contraction is hampered.

[1] See (2), pp. 557.

## 3.1 The two–layer model

We model an isolated fibre of muscle tissue, referred to from now on as the tissue specimen, as a one–dimensional continuum. It may be visualized as a bar of length $l_s$, meant to represent the muscle fibre in its *slack* state, that is, at zero contraction. Moreover, we denote with $l_c$ the contracted and unloaded length of the same fibre, and with $l$ the length that the specimen reveals to an in–vivo observation (the visible length). Here, we limit our analysis to homogeneous deformations and write

$$l_c = \varepsilon_c\, l_s\,, \quad l = \varphi\, l_c = \varphi\, \varepsilon_c\, l_s\,. \tag{3.1}$$

Figure 4 shows the relationship represented by equations (3.1) through a cartoon. Equation $(3.1)_2$ allows a peculiar reading: the *total stretch*

$$\varepsilon := \varphi\, \varepsilon_c \tag{3.2}$$

of the fibre given by the ratio $l/l_s$ is multiplicatively decomposable into an *active stretch* $\varepsilon_c$, measuring the contraction, and a *passive stretch* $\varphi$, responsible for any variation in the elastic energy. Consequently, we require that upon activation $\varepsilon_c < 1$, and under tension $\varphi > 1$; the slack state corresponds to $\varepsilon_c = \varphi = 1$.

This approach is founded on a two–layer kinematic (8), (9), (19), (13). The total stretch $\varepsilon$ lives at the visible layer, insofar it determines the visible length of the specimen. The active stretch (for us, the contraction) $\varepsilon_c$ alters the length of the fibre but does not induce any tension, so that $l_c$ corresponds to a ground state, that is, a state at zero tension. Moreover, it characterizes a hidden layer as it takes up at a macroscopic level phenomena corresponding to microscopic cellular processes.

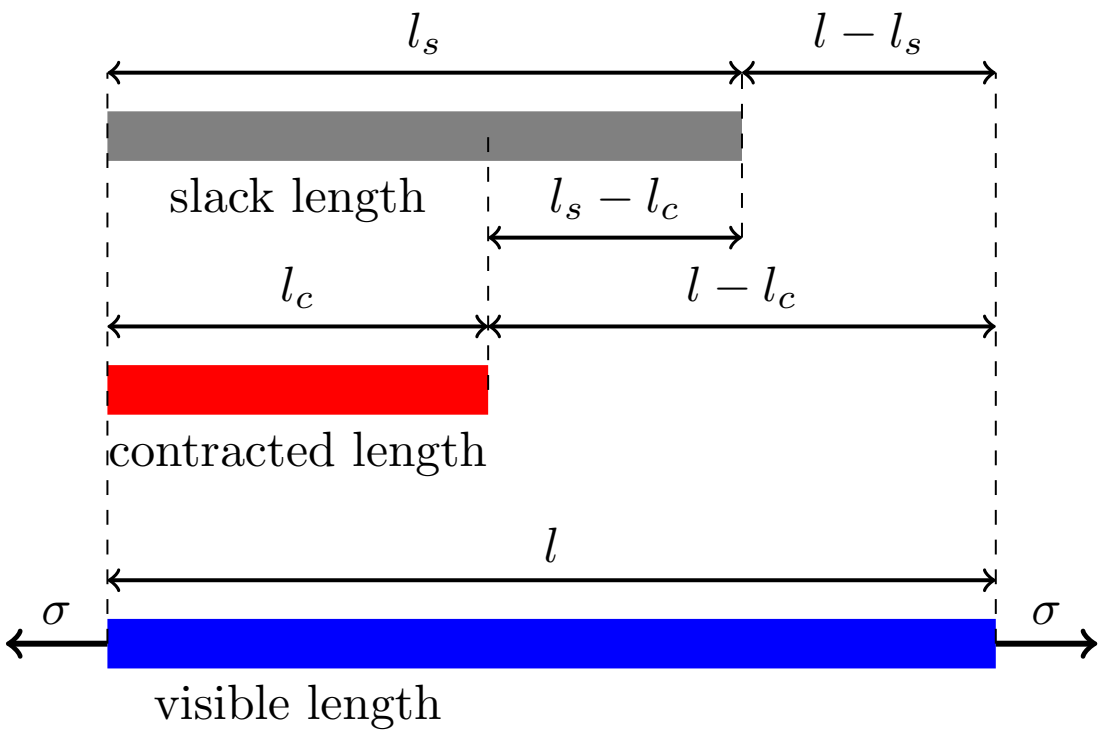

Figure 4: Schematic of muscle contraction. The tension $\sigma$ developed during activation depends on the difference between the visible and the contracted length.

From our point of view, the tension $\sigma$ in the tissue specimen depends on the passive deformation $\varphi$. Nevertheless, the contraction $\varepsilon_c$ alters the tissue response as may be easily shown through the following example. Let $\hat{\sigma}$ be the *constitutive* law relating the passive deformation $\varphi$ to the tension $\sigma$ and let it be represented as

$$\sigma = \hat{\sigma}(\varphi) = Y\, \lambda^3(\varphi)\,, \quad \lambda(\varphi) = \frac{1}{2}(\varphi^2 - 1)\,, \tag{3.3}$$

with $Y$ the elastic stiffness of the tissue specimen. Equation (3.3) is a possible choice for $\hat{\sigma}$ which has the advantage of being simple, and, as shall be shown in the next section, to capture the key features of the EDPVR and ESPVR curves. Nevertheless, different choices for $\hat{\sigma}$ may be done. None of them change the idea underlying the model of muscle tension but just the dependence of the muscle tension on the muscle length. The stretch $\varphi$ is meant to depend on the actual length $l$ and the amount of contraction $\varepsilon_c$ (see equation $(3.1)_2$); $\lambda$ is a common strain measure. For $l = l_s$, $\varepsilon_c = 1$, we have $\varphi = 1$ and thus $\sigma = 0$; it means that the slack length corresponds to a stress–free state.

The rather simple law (3.3) yields a very compliant response for $\lambda \simeq 0$, which becomes quite stiff with increasing values of strain. Thus, with $\varepsilon_c = 1$ it captures, at least qualitatively, the passive response of cardiac muscles, and for $\varepsilon_c < 1$ it describes the response of an activated muscle. To summarize, the notion of contraction that we propose allows us to view any curve tension versus strain as the expression of the same constitutive equation (3.3), which in case of muscle activation ($\varepsilon_c < 1$) stiffens. It is worth noting that equation (3.3) can be used to describe how muscles can control, independently, both position and tension, by allowing a change in $Y$ and/or $\varepsilon_c$.

# 4 Ventricular Pressure-Volume Relationships

Here, we show how the present approach to muscle modeling may be used in the description of the pumping function of the LV. With this aim, we shall look at pressure–volume loops from a novel point of view, and we shall introduce the notion of *contraction–volume loops*.

The macroscopic model we present here is a zero–dimensional model, simple enough to enlighten the key ideas at the basis of the modeling, but capable of capturing the important features of the pumping function of the heart which are collected in the PV loop. Moreover, it is rigorously extendible to the full fledged non-linear 3D elasticity theory. The main mechanical relationships we aim to discuss are not altered by a simplified geometry. This is the reason why, in the following, before turning to complex mathematical models for interrelating ventricular chamber pressure and volume to a suitable measure of contraction, we deal with a spherical approximation of the left ventricle.

For $h$ the constant wall thickness of the surface, a handy relation between the pressure $p$ inside the spherical chamber and the mean tension $\sigma$ generated

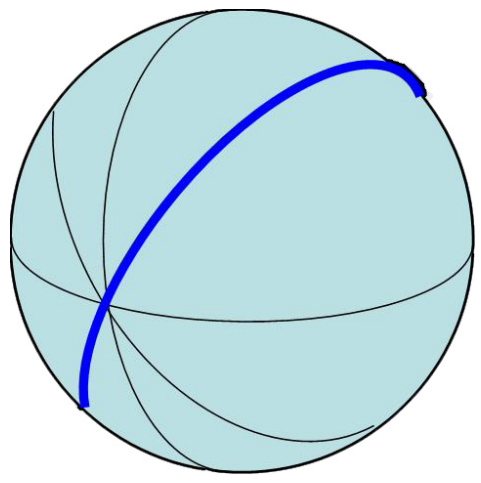

Figure 5: Spherical approximation of the left chamber.

in the LV tissue is the well known Laplace formula:

$$p = 2\,\frac{\sigma\, h}{r}\,. \qquad (4.4)$$

Actually, we are assuming that any hoop fibres in the spherical surface behave as the one–dimensional fibre studied in the previous section. The balance of the spherical surface of radius $r$ is satisfied if equation (4.4) holds. The radius $r$ of the sphere is related to its volume $v$ by

$$v \mapsto r = \hat{r}(v) = \left(\frac{3}{4}\frac{v}{\pi}\right)^{1/3}, \qquad (4.5)$$

so, it is easy to re–write the balance equation (4.4) in terms of pressure and volume instead of pressure and radius. The contracted volume $v_c$ corresponding to the pair $(p, v)$ is a ground volume: it is defined as the pressure–free volume of the sphere which has the volume $v$ under the pressure $p$. In particular, we assume that the pressure–free volume $v_s$ corresponding to the pair $(p_{ed}, v_{ed})$ identifies the slack state of the chamber. Subsequently, we measure contraction from there and assume that the level of muscle activation at the ground volume $v_c$ is:

$$\varepsilon_c = \left(\frac{v_c}{v_s}\right)^{1/3}. \qquad (4.6)$$

It is worth noting that, due to the balance equation (4.4), the pressure–free volume is also a stress–free volume. Nevertheless, it is not contraction–free since the unique pressure–free and contraction–free volume is the slack volume $v_s$. Moreover, we assume that the visible volume $v$ is attained from the contracted volume $v_c$ through an elastic deformation $\varphi$. In the following, we often refer to $r_c$ and $r_s$ as to the radii corresponding to the ground contracted volume $v_c$ and to the slack volume $v_s$, respectively. Of course, $r_c = \hat{r}(v_c)$ and $r_s = \hat{r}(v_s)$; moreover, $\varepsilon_c = r_c/r_s$.

The elastic strain $\lambda$ of the chamber may be written as

$$\lambda = \frac{1}{2}\left( (\frac{r}{r_c})^2 - 1 \right) . \qquad (4.7)$$

Hence, the elastic strain $\lambda$ is zero at the ground volumes ($r = r_c$) which are mechanically relaxed even if places of active deformation (contraction). Moreover, we still write

$$\sigma\, h = \mathbb{Y}\lambda^3 = \mathbb{Y}\left( \frac{1}{2}((\frac{r}{r_c})^2 - 1) \right)^3 , \quad \mathbb{Y} = Y\, h , \qquad (4.8)$$

for the tension developed into the chamber whose slack radius is $r_s$ and denote with $\mathbb{Y}$ the elastic membrane stiffness of the chamber. Equations (4.4), (4.7), and $(4.8)_1$ turn out a basic equation relating the pressure $p$ and the volume $v$ of a spherical surface characterized by the slack radius $r_s$ and by the stiffness $\mathbb{Y}$ when contraction attains the value $\varepsilon_c$:

$$p = 2\,\frac{\mathbb{Y}}{r}\left( \frac{1}{2}((\frac{r}{r_s}\varepsilon_c^{-1})^2 - 1) \right)^3 , \quad r = \hat{r}(v) . \qquad (4.9)$$

Let us note that, fixed $r_s$, equation (4.9) can be viewed as a function $f$ relating pressure, volume and contraction, with stiffness $\mathbb{Y}$ acting as a parameter:

$$f(p, v, \varepsilon_c; \mathbb{Y}) = 0 . \qquad (4.10)$$

The function $f$ can be solved with respect to each one of the three variables $(p, v, \varepsilon_c)$, and completely characterizes the pumping action of the heart. In particular, as it will be shown, it comprises the key curves EDPVR and ESPVR. Equation $f$, solved with respect to $p$, gives a pressure–volume relationship depending on the contraction

$$p = p(v, \varepsilon_c; \mathbb{Y}) . \qquad (4.11)$$

Specifically, $p = p(v, 1; \mathbb{Y})$ yields the EDPVR curve, and $p = p(v, \varepsilon_{c,es}; \mathbb{Y})$, with $\varepsilon_{c,es}$ the maximum value attained by $\varepsilon_c$, the ESPVR curve. Finally, solving $f$ with respect to $\varepsilon_c$, gives the amount of contraction corresponding to any specific pair $(p,\, v)$

$$\varepsilon_c = \varepsilon_c(p, v; \mathbb{Y}) . \qquad (4.12)$$

Thus, the role of (4.10) is analogous to that of (2.1) and of the other empirical formulas relating pressure and volume (20), that is, to subsume through macroscopic quantities the main features of PV loops. Nevertheless, it replaces the notion of time–varying elastance $E(t)$ with that of time–varying contraction $\varepsilon_c(t)$, thus expressing through a physically based parameter the subtle notion of chamber contraction. In the end, let us note that the assumption that the stiffness $\mathbb{Y}$ does not change during contraction is just a simplifying hypothesis which, however, does not alter the capacity of the model and may be easily revised.

## 4.1 A case study

We discuss our point of view with reference to a specific pressure–volume loop extracted by (15). There, with reference to a normal human patient, the pairs pressure(mmHg)–volume(ml) are measured and the PV loop shown in figure 6 is generated. Precisely, points from 1 to 2 describe the isovolumic contraction; points from 2 to 3 describe the ejection phase; points from 3 to 4 describe the isovolumic relaxation; and points from 4 to 1 describe the filling phase. For a better comprehension, in figure 7, top, we represent with different colours the time course of the pressure corresponding to the four characteristic phases of the cardiac cycle: blue for isovolumic contraction, red for ejection, green for isovolumic relaxation, and black for ventricular filling. We assume that the points labelled 1 and 3 correspond to the end–diastolic and the end–systolic pressure–volume pairs, respectively. When fixing the value of the elastic stiffness $\mathbb{Y}$ of the chamber to 1500 mmHg cm ((21)) and defined $r_s$ as the radius corresponding to the volume characteristic of the point 1, we recover the contraction measure associated to every $(p,\, v)$ state in the loop. Denoted with $(\varepsilon_c)_i$ the contraction corresponding to the state identified by the pair $(p_i, v_i)$, from equation (4.12), we find

$$(\varepsilon_c)_i = \varepsilon_c(p_i, v_i; \mathbb{Y}) .$$

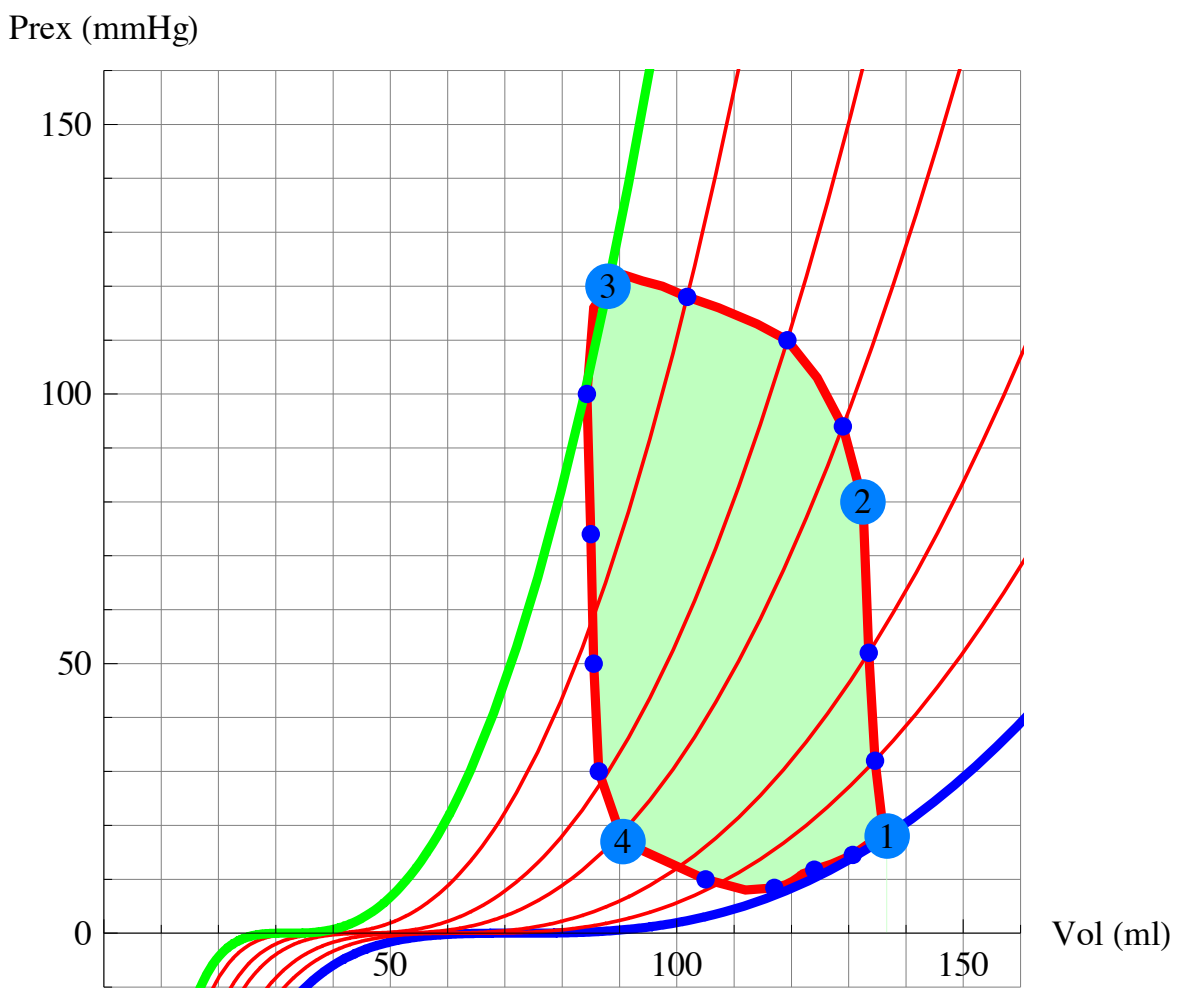

Figure 6: A typical PV loop of a normal human patient, as measured in (15), with our ESPVR and EDPVR curves superimposed, as equation (4.9) dictates; the large blue dots correspond to the same four key points shown in figure 1.

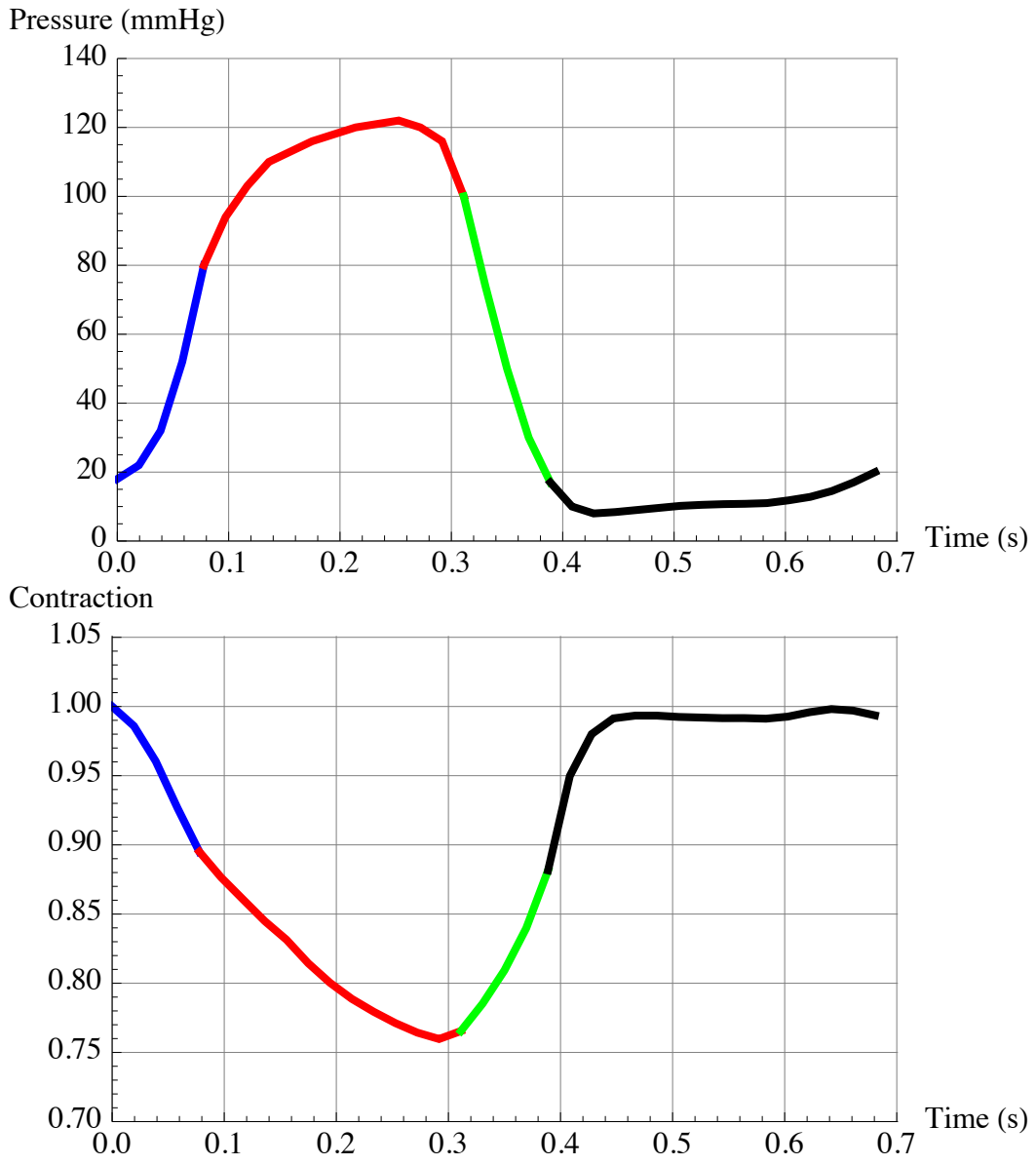

Figure 7: A typical pressure cycle VS time (top, measured from (15)) and the corresponding contraction (bottom) as given by equation (4.9).

In figure 7, bottom, the time course of the contraction cycle is represented: the end–diastolic state corresponds to $\varepsilon_c = 1$ and the end–systolic state to $\varepsilon_{c,es} = 0.766$. Corresponding to these values, equations (4.9) gives the EDPVR and ESPVR curves which are represented in figure 6 (blu and green solid line, respectively) as superimposed on the PV loop extracted by (15). Moreover, the transition from EDPVR to ESPVR may be derived through a generalization to any point during the cardiac cycle of the procedure used to extract the EDPVR and the ESPVR relationships from equation (4.9). The pressure–volume curves corresponding to a few intermediate points in the cycle are shown in figure 6 (red lines). Let us note the following key points.

- The EDPVR curve is intrinsically nonlinear, as it is expected and as equation (4.9) rules. Moreover, it shows that, at subphysiological volume ranges, increasingly negative pressures are required to reduce volume. Nevertheless the simplicity of the theoretical model, a likely (see (1)) effect is caught by the model.
- The volume axis intercept $v_s$ of the EDPVR curve corresponds to the slack chamber at zero stress which is determined at a ventricular pressure of 0 mmHg. As noted in (1), this intercept differs from the volume axis intercept of the ESPVR which represents the volume of the chamber at zero stress and at maximum contraction.
- According to the experiments, (see figure 5 in (1)), figure 6 shows a smooth transition from the EDPVR to ESPVR curves corresponding to a stiffening of the chamber due to the muscle activation.
- Actually, the volume axis intercepts of the pressure–volume curves at different values of contraction differ one from the other. In other words, the volume axis intercept is not totally independent of the contraction state (2).

- Figure 6 shows a non linear ESPVR. Indeed, the ESPVR is in general non linear even if reasonably linear relationships are commonly used to characterize properties of the chamber with muscles in a state of activation (see equation (2.1)).

As well as a new point of view in looking at pressure–volume loops, our model allows the association of any PV loop with a contraction–volume loop. It captures at a glance a hidden property of the left chamber and gives an idea of the contractility of the left ventricle around all the key phases of the cycle. Precisely, figure 8 shows the contraction–volume loop corresponding to the numeric PV loop described in figure 6. It shows some key interesting features such as a variability of the contraction measure in a range which is in agreement with scientific literature (2) and a strong similarity to the experimental loop in figure 3.

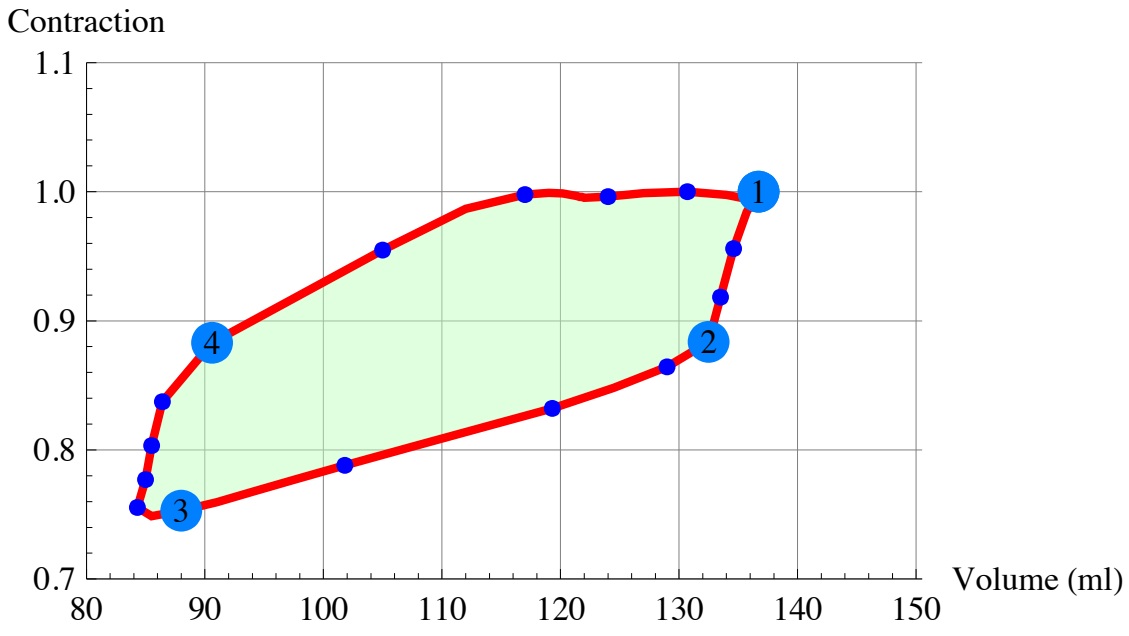

Figure 8: Contraction along the cardiac cycle as function of volume.

The following cartoon sketches efficiently our point of view. Blue balls aim to represent the visible states of the left chamber in the four key points of the cycle. Correspondingly, red balls represent the contracted and stress–free states. As an example, let us follow what happens from point 1 to point 2. During the isovolumic contraction, the visible volume of the balls is roughly the same (136ml versus 132ml), that is, red balls 1 and 2 have almost the same radius. At the same time, at a hidden layer something is occurring and the red ball 2 is smaller than the red ball 1 (identifying the slack state of the chamber) because a contraction is acting and making the difference between states 1 and 2.

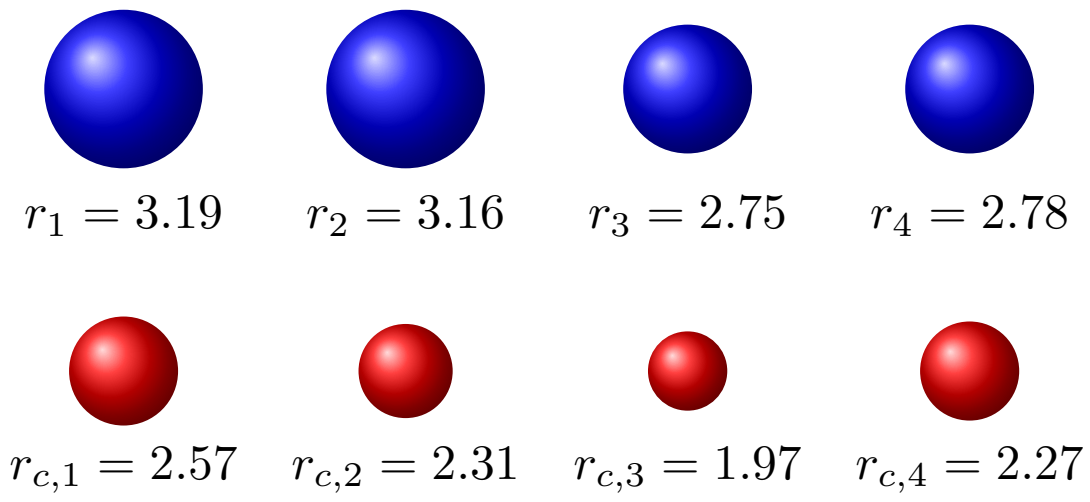

Figure 9: Schematic of LV contraction. The top row shows the visible state of the LV in the four key points of the cycle; the bottom row shows the corresponding contracted, but unloaded, LV (actually, the first point represents the slack state). All the measures are in cm.

Finally, let us introduce the deformation $\bar{\varepsilon}$ with respect to the end–diastolic state; the corresponding strains $(\bar{\varepsilon}^2 - 1)/2$ along the cardiac cycle are shown in figure 10 and, interestingly, are in agreement with (22) where circumferential strains at different points of the left ventricle are measured.

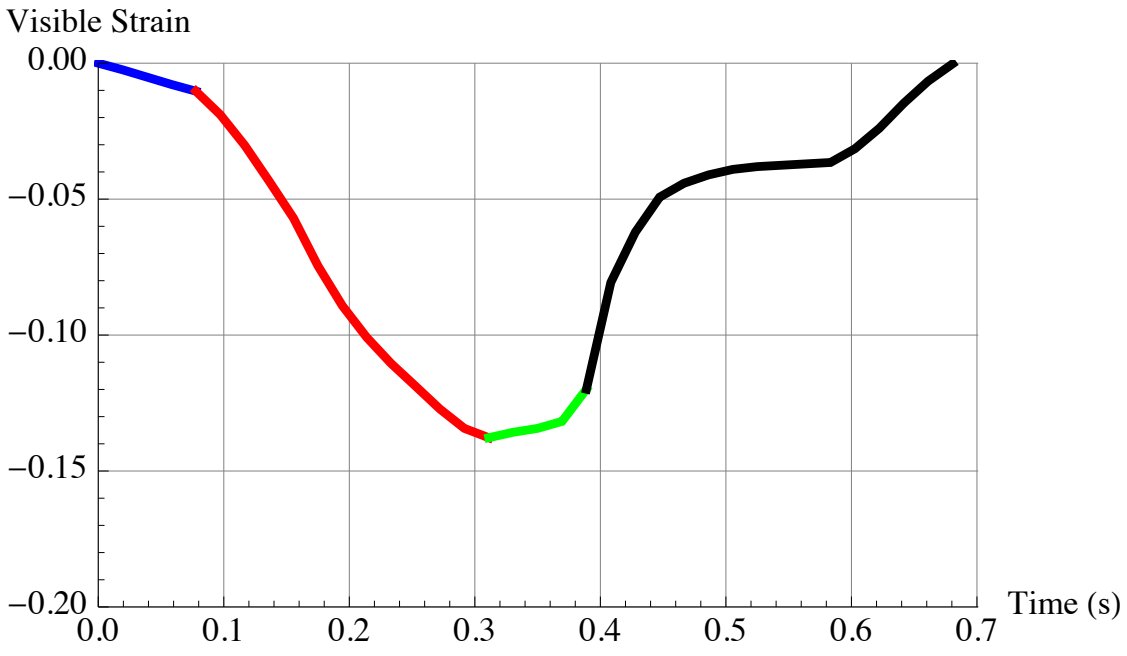

Figure 10: Visible strains vs time along the cardiac cycle.

## 4.2 Ventricular–vascular coupling: effects of preload and afterload

In the previous sections, we presented our interpretative model of pressure–volume loops based on a distinguished notion of contraction.

Here, we show further appealing characteristics of our modeling, that is, its ability to predict some alterations of the PV loop and the contraction loop, as consequences of changes in preload and/or afterload conditions.

As explained in (23), "if an intervention is performed that acutely *changes the loading conditions* on the heart but has *no effect on myocardial contractility* (e.g., transient inferior vena caval occlusion to reduce preload, administration of phenylephrine to increase afterload, etc.) a *family of loops* is obtained. The end-systolic and end-diastolic points of these loops delineate two distinct boundaries".

Let us observe that the shape of the loops is the same for every element of the family, in such a way that every loop can be described simply through the segment ED–ES, moving between the two rails EDPVR and ESPVR. (see figure 11). Further, we assume that all the other points of the loop are scaled to follow the transformation of the segment ED–ES. Then, by choosing a PV loop as a reference, we can reasonably describe the whole loop family through a two–parameter stretch transformation group. From a geometrical point of view, the simplest choice of the two key parameters is the arch-length (say $(s_1, s_2)$) on EDPVR and ESPVR. On the other hand, from a physical point of view better choices are the pairs $(p_{ed}, p_{es})$, $(v_{ed}, v_{es})$, and $(p_{gn}, v_{str})$, with the pressure gain $p_{gn} := p_{es} - p_{ed}$. Actually, we introduce a further parameter which quantifies the properties of the arterial system, the *effective arterial elastance* $E_a$ defined by

$$E_a = \frac{p_{es} - p_{ed}}{v_{ed} - v_{es}} = \frac{p_{gn}}{v_{str}} \,. \qquad (4.13)$$

The importance of $E_a$ is twofold: it admits a straightforward graphical representation on the PV diagram which highlights the role of the two key points 1 (ED) and 3 (ES), see Fig. (11). In fact, $E_a$ represents the slope of the segment ED–ES and is strictly related to the *total peripheral resistance* $R_p$, defined as the ratio between the pressure gain $p_{gn}$ and the cardiac output $(v_{ed} - v_{es})\, H_R$, with $H_R$ denoting the heart rate:

$$R_p = \frac{p_{es} - p_{ed}}{(v_{ed} - v_{es})\, H_R} = \frac{p_{gn}}{v_{str}\, H_R} = \frac{E_a}{H_R} \,. \qquad (4.14)$$

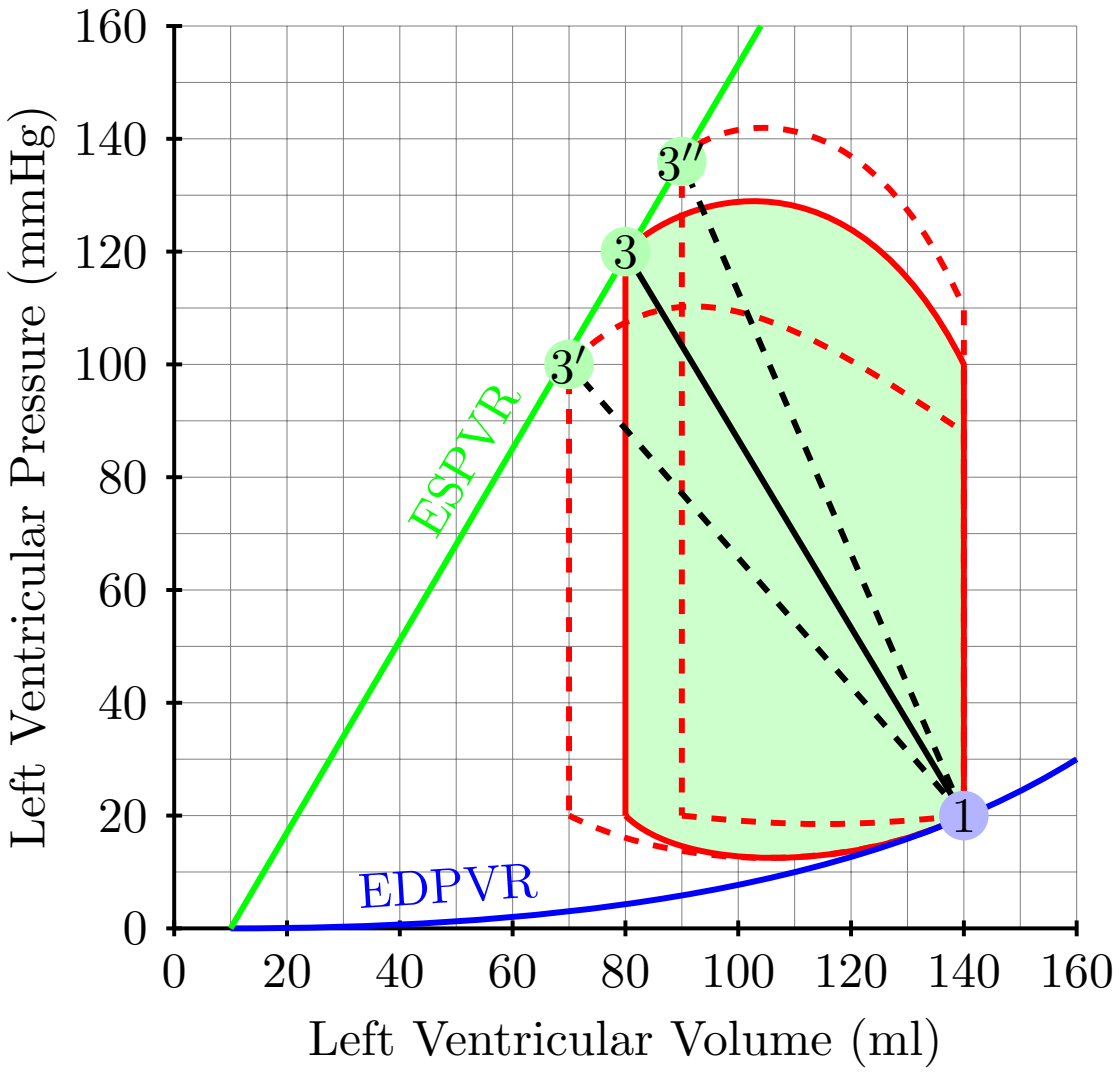

Figure 11: The effective arterial elastance $E_a$ may be represented in the PV diagram by the line connecting point 1 to 3, that is, ED to ES (the slope of this line is actually $-E_a$). As example, any alteration of the afterload moves point 3 along the ESPVR, thus changing $E_a$; the new PV loop is related to the baseline by an affine transformation (baseline is represented with solid line, alterations with dashed lines).

Equation (4.14) quantifies the ventricular-vascular coupling, as it relates heart rate, stroke volume, and peripheral resistance. For example, it shows that altered peripheral resistance results in a variation of the heart rate and/or the effective arterial elastance, that is, of the points ED, ES. As shown in (1) a useful parametrization of the curve family can be $(v_{ed}, E_a)$, in the sense that we can choose $(v_{ed}, E_a)$ and obtain, for example,$v_{es}$.

In the following, with the aim of showing the typical alterations of the PV loops due to changes in preload and/or afterload conditions, we will use the transformation group based on the two parameters $(p_{ed}, p_{es})$. Thus, the altered PV loop may be reconstructed as follows: let us consider pressure as independent variable, and the function (4.10) solved with respect to the volume: $v = v(p, \varepsilon_c; Y)$; changes in preload and/or afterload implies that $p_{ed} \mapsto \bar{p}_{ed} = p_{ed} + \triangle p_{ed}$ and/or $p_{es} \mapsto \bar{p}_{es} = p_{es} + \triangle p_{es}$, respectively; accordingly, volumes $v_{ed}$ and/or $v_{es}$ change as follows

$$\begin{aligned} v_{ed} &\mapsto \bar{v}_{ed} = v(\bar{p}_{ed}, \varepsilon_{c,ed}; \mathbb{Y})\,, \\ v_{es} &\mapsto \bar{v}_{es} = v(\bar{p}_{es}, \varepsilon_{c,es}; \mathbb{Y})\,. \end{aligned} \tag{4.15}$$

Thus, under the assumption that $\varepsilon_{c,ed}$, $\varepsilon_{c,es}$, and $\mathbb{Y}$ stay constant, points ED and/or ES move along the EDPVR and/or ESPVR curves, respectively. Moreover, any other point $(\bar{p}_i, \bar{v}_i)$ in the altered PV loop may be given in terms of the baseline points $(p_i, v_i)$ through the affine transformation

$$\begin{aligned} \bar{p}_i - \bar{p}_{ed} &= \bar{p}_{gn}/p_{gn}\,(p_i - p_{ed})\,, \\ \bar{v}_i - \bar{v}_{ed} &= \bar{v}_{str}/v_{str}\,(v_i - v_{ed})\,. \end{aligned} \tag{4.16}$$

Eventually, equation (4.10), solved with respect to contraction, $\varepsilon_c = \varepsilon_c(p, v; Y)$, yields the new values of contraction $\bar{\varepsilon}_{ci}$ needed to sustain the altered conditions

$$\bar{\varepsilon}_{ci} = \varepsilon_c(\bar{p}_i, \bar{v}_i; \mathbb{Y})\,. \tag{4.17}$$

These values, a prediction of the model, may shed light on the amount of the physiological compensation needed as the consequence of altered preload and/or afterload: in order to pump blood effectively under the new conditions, that is, to fulfill the condition that PV loops transform affinely, muscle contractions have to change.

Figure (12) shows the effects of altered preload on the baseline PV loop shown in figure 6; next figure (13) shows the associated alterations in the contraction loop as predicted by (4.17). Figure (14) and (15) show the same effects for altered afterload.

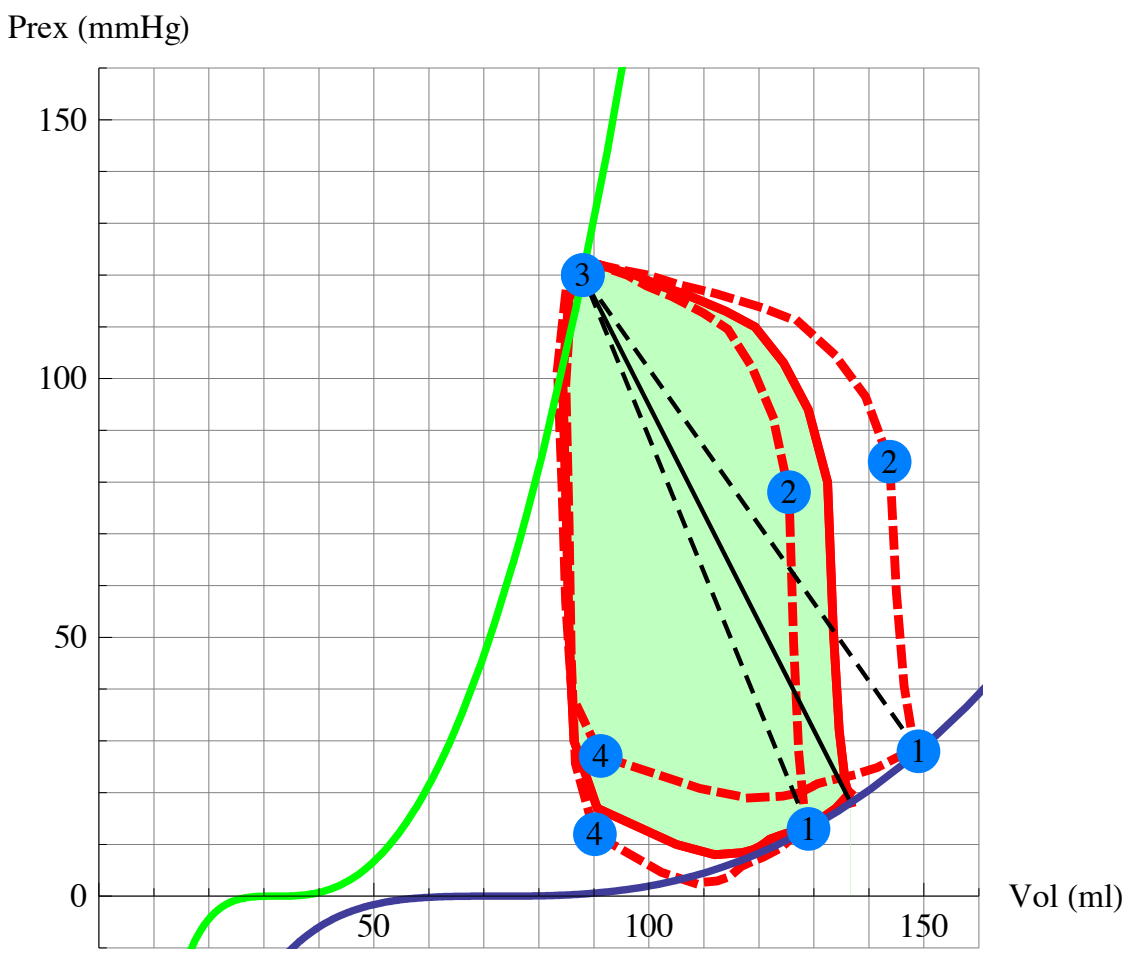

Figure 12: Changes in the baseline PV loop (red line, solid) due to an increase and a decrease of the preload (black line, dashed).

## 5 Geometrical and material remodeling at a glance

In the previous sections, we presented our interpretative model of pressure–volume loops based on a distinguished notion of contraction which allows a macroscopic modeling of the pumping action of the heart. Pressure–volume loops are viewed as determined by three main state variables, pressure, volume, and contraction related through equation (4.10). Actually, other variables determine the characteristics of PV loops. A few of them depend on the circulatory system and are represented in the pressure–volume plane by the notion of preload and afterload discussed in the last section. Other variables depend on the heart structure; we think of the size and the stiffness of the left chamber and refer to the ability of the heart tissue to shorten and relax. All these elements strongly vary in the presence of heart failure which, in its turn, may be revealed just through the analysis of the corresponding pressure–volume loops.

Here, we aim to discuss as an independent variation of the size, the stiffness, and the ability to contract of the left chamber, represented within our model through the triplet $(v_s, \mathbb{Y}, \varepsilon_c)$, allows to recover some typical alterations which pressure–volume loops show in presence of heart failure. Even if applied here in a simple context, the basic idea is that a well–founded model of the pumping function of the heart allows the intervention on specific parameters and gives back meaningful information. So, in a patient–specific analysis in the presence of altered pressure–volume loops, the model may suggest the role of different parameters in determining the alteration and the right tool for recovering the normal situation.

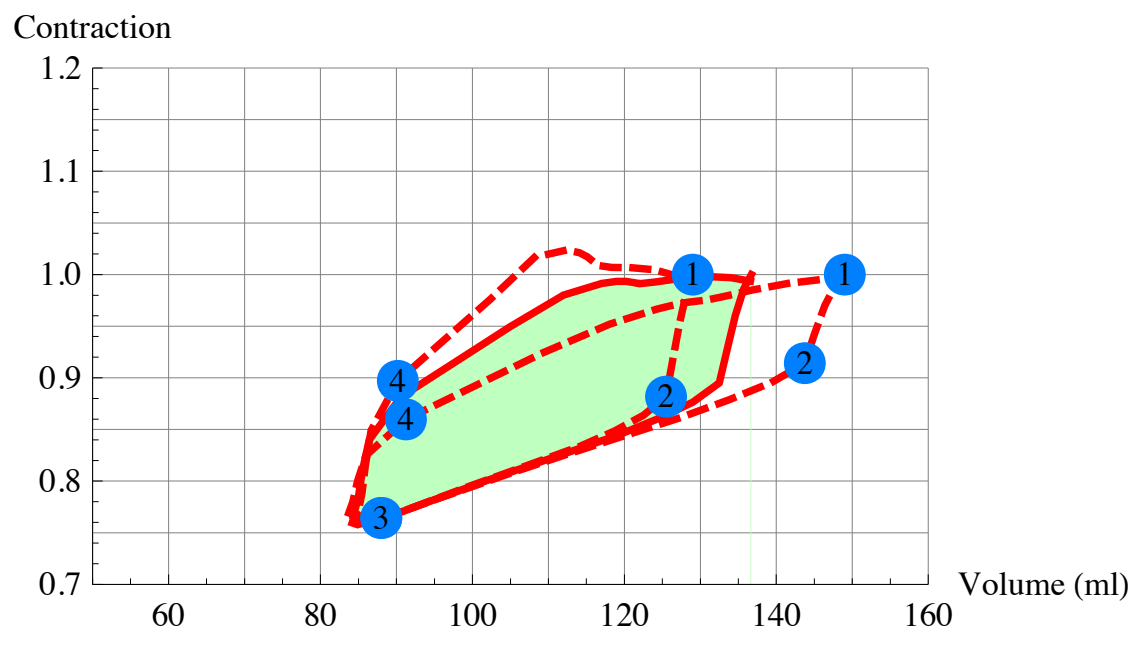

Figure 13: Alterations in the baseline contraction–volume loop (red line, solid) corresponding to the increase and decrease of the preload (blue line, dashed).

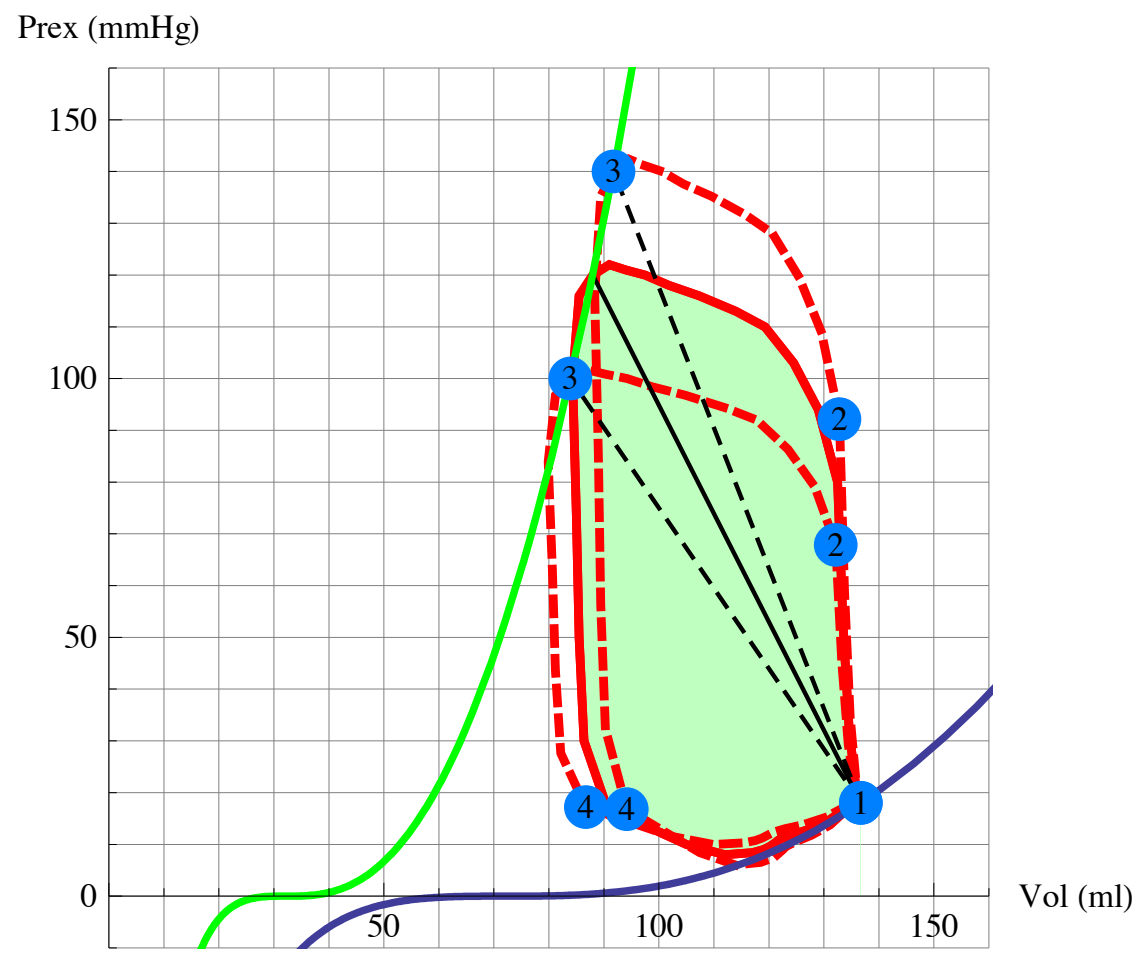

Figure 14: Changes in the baseline PV loop (red line, solid) due to an increase and a decrease of the afterload (blue line, dashed).

## Change in size

We denoted with $v_s$ the volume axis intercept of the pressure–volume curve corresponding to the end–diastolic state which views the chamber as relaxed and stress–free. Moreover, we remember that,in our modeling, $v_s$ is the reference volume for the measure of contraction as equation (4.6) says. We model size abnormalities just as alteration of the relaxed and stress–free size of the chamber through the scalar remodeling parameter $\alpha_m$ such that $\alpha_m\, r_s$ measures the radius of the altered chamber. For $\alpha_m < 1$ and $\alpha_m > 1$ a reduction and an enlargement of the left chamber is accounted for, respectively. So, let us write equation (4.10) as

$$f(p, v, \varepsilon_c(r_s); \mathbb{Y}) = 0\,, \qquad (5.18)$$

in such a way to make evident the role of the radius $r_s$ as reference radius for the contraction measure. Equation (5.18), solved with respect to $v$ gives back the volumes $v_i$ corresponding to the pressures $p_i$: $v_i = v(p_i, \varepsilon_c(r_s); \mathbb{Y})$. In the presence of a size alteration, we have a new reference radius $\alpha_m\, r_s$ and, corresponding to the same pressures, new volumes $v_i^*$ as

$$v_i^* = v(p_i, \varepsilon_c(\alpha_m\, r_s); \mathbb{Y})\,. \qquad (5.19)$$

The altered PV loop (dashed lines) corresponding to the old pressures and to the remodeling parameter $\alpha_m = 1.04$ is shown in figure 16 together with the normal PV loop (solid lines). The changes in EDPVR and ESPVR show as the enlarged chamber induces a decreased cardiac efficiency. Indeed, the change in the EPVR curve says that at the same pressure a larger volume of the chamber is attained. Moreover, an increase of the stroke volume turns out when enlargement occurs while pressures stay normal.

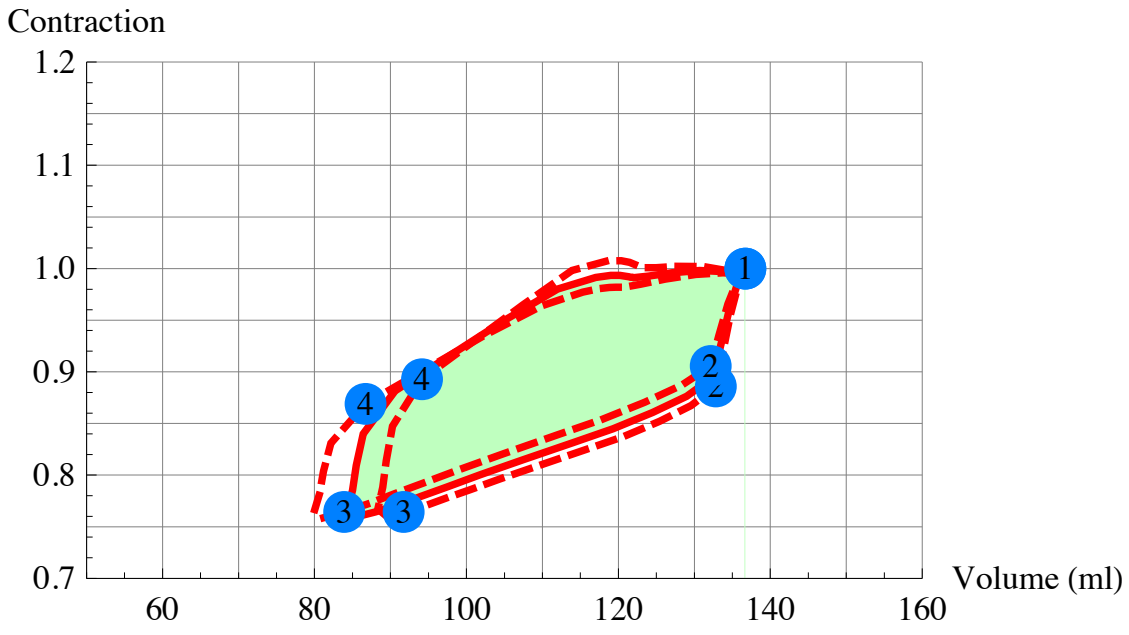

Figure 15: Visible strain vs time along the cardiac cycle.

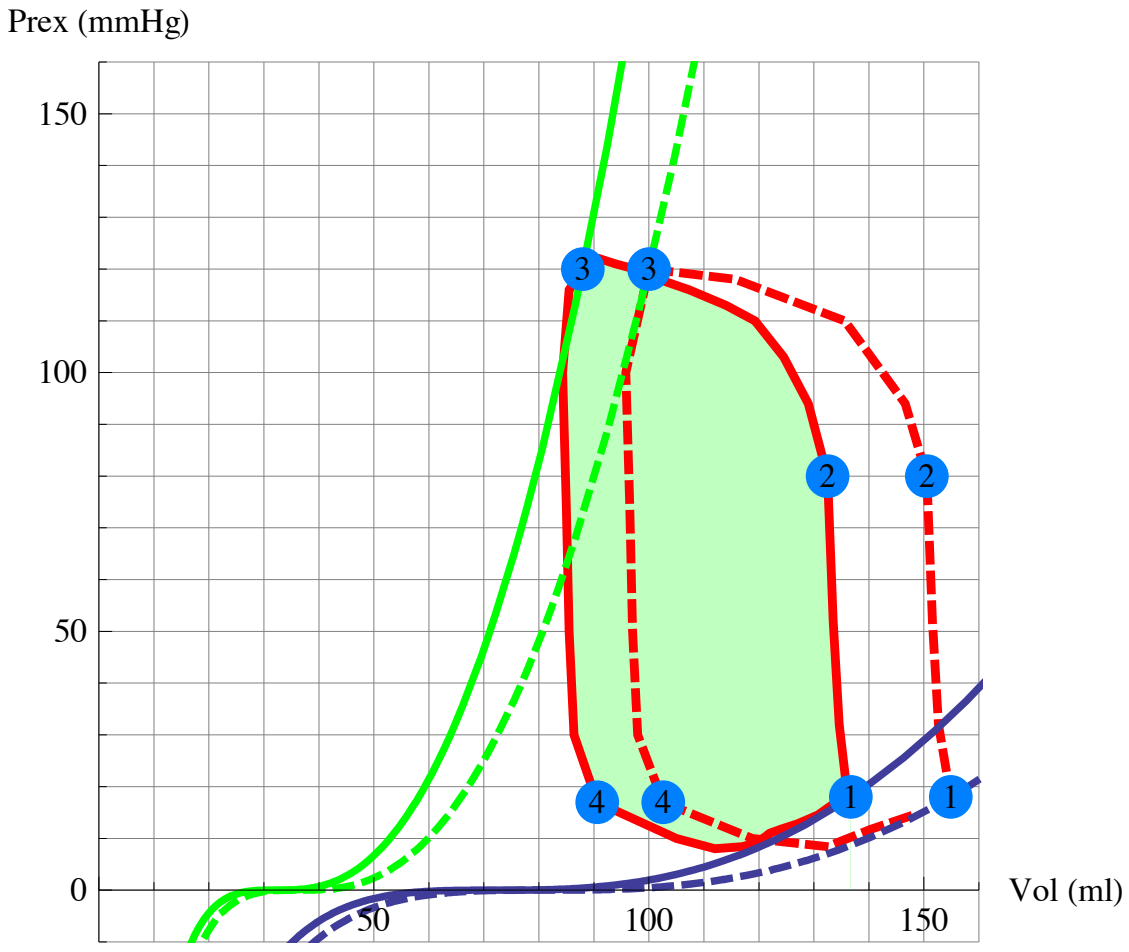

Figure 16: The altered PV loop in presence of remodeling ($\alpha_m > 1$ (dashed lines) versus the normal loop (solid lines).

#### 5.0.1 Change in contractility

The ability to contract and relax of the left chamber is accounted for through the parameter $\varepsilon_c$. We note that equation (4.12) allows any pressure–volume pair in a PV loop to associate with the corresponding contraction $\varepsilon_c$. Hence, a variation of $\varepsilon_c$ means a variation of the ability of the chamber to contract at any time along the cardiac cycle. Of course, it embodies a variation of the maximum contraction $\varepsilon_{c,es}$ and, consequently, a variation of the ESPVR curve. We account for such a variation through the scalar contractility parameter $\alpha_c$ so that

$$\varepsilon_c + \alpha_c\,(1 - \varepsilon_c) \tag{5.20}$$

measures the altered ability to contract. Still, equation (4.10), solved with respect to $v$ gives back the new volumes corresponding to the normal pressures when the contraction of the chamber is uniformly altered:

$$v_i^* = v(p_i, \varepsilon_c + \alpha_c\,(\varepsilon_c - 1); \mathbb{Y})\,. \tag{5.21}$$

For $\alpha_c < 0$ and $\alpha_c > 0$ a decrease and an increase in $\varepsilon_c$ is produced, respectively. Figure 17 shows as an increase in contractility corresponding to $\alpha_c = 0.4$ is revealed by the stiffening of the ESPVR curve, which is related to the contraction measure at the point 3. Of course, the EDPVR curve stays unchanged as it corresponds to a contraction measure $\varepsilon_c = 1$.

#### 5.0.2 Change in stiffness

In conclusion, we discuss the role of the stiffness of the chamber in the modeling through the analysis of the consequences of a variation in the chamber stiffness. Firstly, let us note that here $Y$ is a global measure of stiffness and it does not allow to account for the stiffening of portions of the left ventricle due to the replacement of the normal tissue by scar such as after a large myocardial infarction. Nevertheless, the typical degradation of the elastic properties of the chamber occurring in heart failure may be accounted for through uniform decrease of the chamber stiffness $\mathbb{Y}$. Equation (4.10), solved with respect to $v$ gives back the volumes corresponding to an altered stiffness of the chamber when pressures and contractions stay unchanged:

$$v_i^* = v(p_i, \varepsilon_c; \alpha_s\,\mathbb{Y})\,, \quad \alpha_s > 0\,; \tag{5.22}$$

$\alpha_s$ is the key element in the parametric analysis. Figure 18 shows as pressure–volume loops change with respect to the normal one when a strong decrease in the chamber stiffness is accounted for.

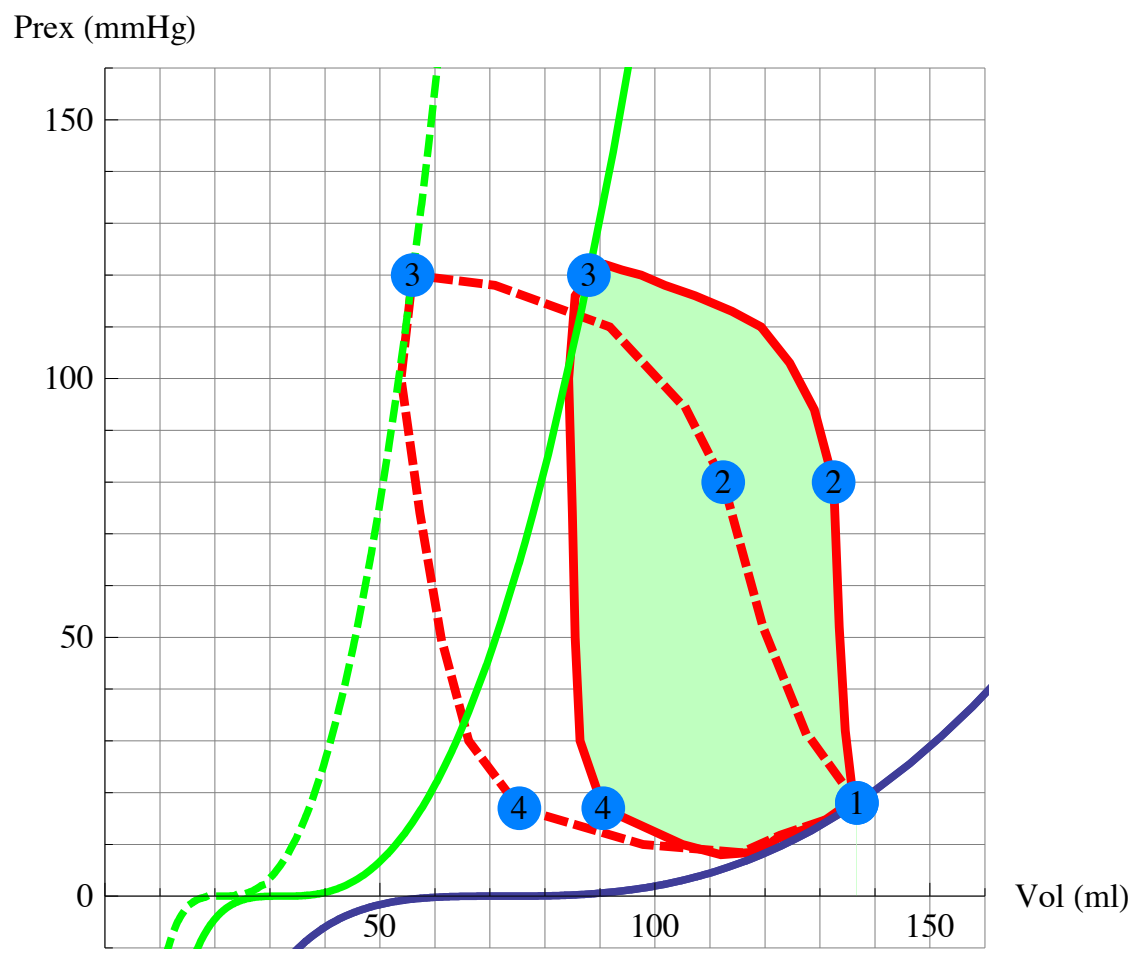

Figure 17: The altered PV loop in presence of a positive variation in contractility ($\alpha_c > 0$ (dashed lines) versus the normal loop (solid lines).

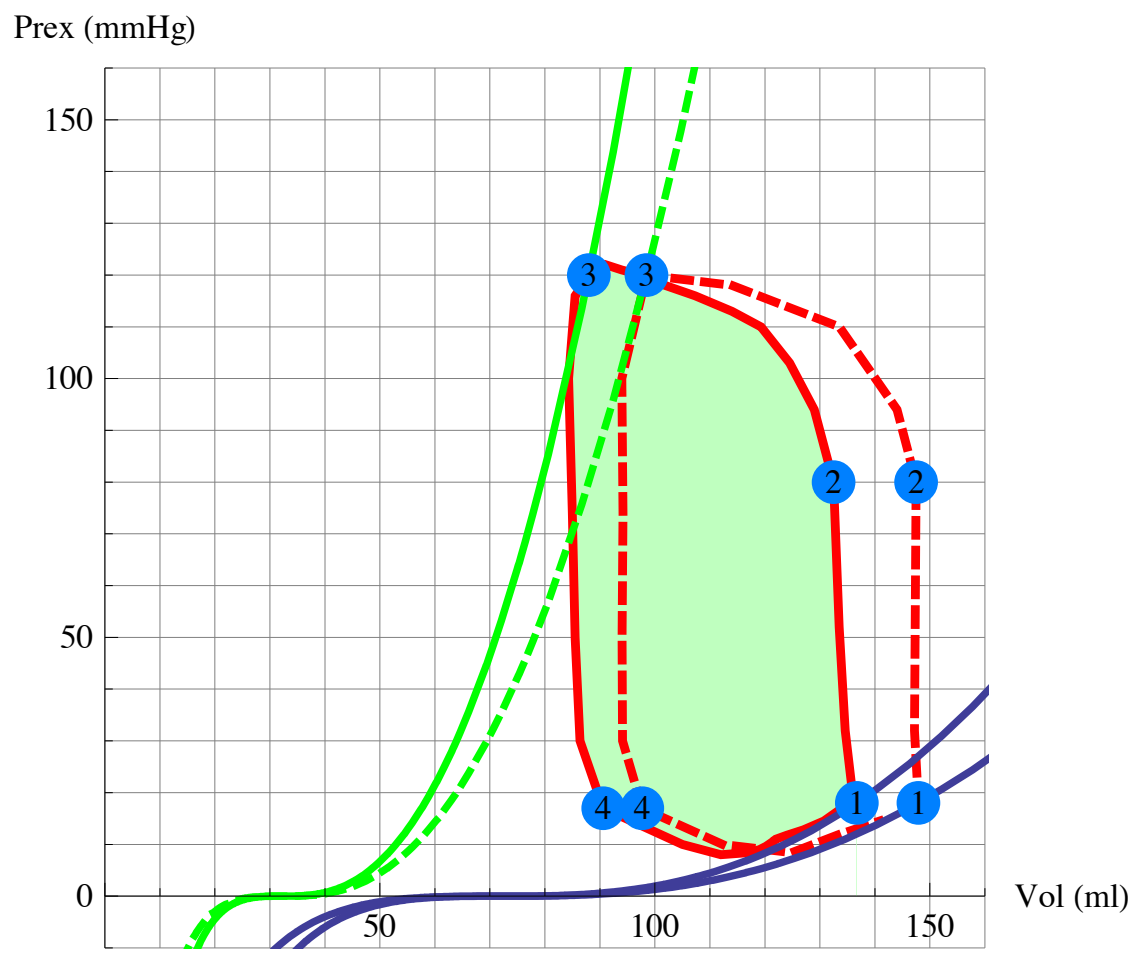

Figure 18: The altered PV loop in presence of a decrease in the chamber stiffness ($\alpha_s < 1$ (dashed lines) versus the normal loop (solid lines).

# 6 Conclusion

A novel point of view is introduced in the modeling of the activable nature of cardiac tissue defining the muscle contraction as an active deformation of the tissue. Here, with reference to a simple heart model, a mechanical interpretation of cardiac PV loops is proposed based on a special modeling of muscle tissue.

In our opinion, the simplicity of the model helps to enlighten the basic characteristics of the PV loop and, as will be shown in future works, to discuss typical heart dysfunctions detectable through the PV loop. Nevertheless, a less simple heart modeling would be more appropriate to account for the complex material structure of the walls of the left chamber as well as for the interaction between the mechanics and the electrophysiology of the cardiac tissue. In these directions, further work must be done following along the lines already identified in (13), (14), (12).